\documentclass[12pt, a4paper]{article}

\usepackage{lipsum}
\usepackage{graphicx}
\usepackage{cite}
\usepackage{booktabs}
\usepackage{caption}
\usepackage{subcaption}
\usepackage{hyperref}
\usepackage{authblk}
\usepackage{float}
\usepackage[margin=1in]{geometry}
\usepackage{fancyvrb}
\usepackage{breqn}
\usepackage{hyperref}

\usepackage[multiuser,nomargin,inline,index]{fixme}

\graphicspath{{./images/}}

\author[1]{Josh Collyer}
\author[1,2]{Tim Watson}
\author[1]{Iain Phillips}
\affil[1]{Computer Science, Loughborough University}
\affil[2]{The Alan Turing Institute}

\title{FASER: Binary Code Similarity Search through the use of Intermediate Representations}

\begin{document}

\maketitle

\begin{abstract}
\noindent
Being able to identify functions of interest in cross-architecture software is
useful whether you are analysing for malware, securing the software supply chain
or conducting vulnerability research. Cross-Architecture Binary Code Similarity
Search has been explored in numerous studies and has used a wide range of
different data sources to achieve its goals. The data sources typically used
draw on common structures derived from binaries such as function control flow
graphs or binary level call graphs, the output of the disassembly process or the
outputs of a dynamic analysis approach.  One data source which has received less
attention is binary intermediate representations. Binary Intermediate
representations possess two interesting properties: they are cross architecture
by their very nature and encode the semantics of a function explicitly to
support downstream usage. Within this paper we propose Function as a String
Encoded Representation (FASER) which combines long document transformers with
the use of intermediate representations to create a model capable of cross
architecture function search without the need for manual feature engineering,
pre-training or a dynamic analysis step. We compare our approach against a
series of baseline approaches for two tasks; A general function search task and
a targeted vulnerability search task. Our approach demonstrates strong
performance across both tasks, performing better than all baseline approaches. 
\end{abstract}

\section{Introduction}

Binary Code Similarity Search aims to provide a means of finding compiled
functions which are similar to a given query function. Being able to achieve
this is useful when wanting to identify similarities between malware
functionality, identify function re-use or to understand whether a piece of
software contains known vulnerabilities. This is a complex undertaking. Factors
ranging from the diversity of toolchains to compiler optimization options mean
that functions can be represented differently across binaries.  The diversity of
ISAs is vast when viewing the problem through an embedded computing lens where
software can be used within systems ranging from a MIPS-based embedded 5G modem
to a 1750A-based subsystem in a US Apache Helicopter.

This problem is not new, however, and Binary Code Similarity Search has been
tackled using a range of different methods. In particular, natural language
processing (NLP) approaches have been transitioned from other domains and
applied to binary analysis tasks. Early approaches such as
SAFE~\cite{massarelli2019safe}, asm2vec~\cite{ding2019asm2vec} and
InnerEye~\cite{zuo2018neural} explored using NLP for binary code search
utilising state-of-the-art approaches. The literature then developed and moved
onto explore using the advances in Transformer architectures in approaches such
as jTrans~\cite{wang2022jtrans}, PalmTree~\cite{li2021palmtree} and
Trex~\cite{pei2020trex}, all of which use a similar pre-training methodology as
BERT~\cite{devlin2018bert} with the addition of domain specific tasks or binary
analysis specific data sources.

The aforementioned approaches all suffer from challenges with constructing a
vocabulary which is able to cover the range of possible inputs. This is referred
to as the Out of Vocabulary (OOV) problem~\cite{li2021palmtree} and stems from
the use of assembly instructions as input which can include a broad range of
possible values such as memory addresses and opcodes. Even after normalization
the number of possible inputs continues to increase with the number of supported
architectures due to implementation-specific nuances. In order to overcome this
challenge, some approaches have instead sought to use an intermediate
representation (IR) as the input data format. For example,
XLIR~\cite{gui2022cross} uses the LLVM IR to conduct binary to source function
search and Penwy~\cite{pewny2015cross} uses the VEX IR alongside a form of
concolic execution to create \emph{bug signatures} for known bugs to conduct
vulnerability search. Neither of these approaches however tackles binary
function search directly using only the IR without any additional inputs.

Within this paper, we propose Function As a String Encoded Representation
(FASER)\footnote{https://github.com/br0kej/FASER} which combines the long
document transformer architecture, Longformer~\cite{beltagy2020longformer}, with
the use of radare2's~\cite{radare2} Evaluable String Intermediate Language
(ESIL) to create a cross architecture model which is capable of binary function
search across multiple different architectures. Through using an IR as the input
data type, we side step the issue of having to normalize for each assembly
language and instead normalize once across a single common representation.

The key contributions of this paper are:
\begin{enumerate}
	\item A binary function representation as IR \emph{Functions as Strings}
	which requires no additional data processing effort other than
	normalization.
	\item A cross-architecture model which combines the usage of IRs alongside
	longer context transformers and demonstrate its usefulness for
	cross-architecture function search and known vulnerability detection.
	\item We demonstrate that it's possible to get strong cross-architecture
	binary search performance using a transformer architecture without the need
	for pre-training and instead using deep metric learning to train directly
	for the binary function search objective.
	\item We conduct, as far as the authors are aware, the first experiment for
	cross-architecture function search using RISC-V architecture as part of the
	experimental methodology.
\end{enumerate}

This paper is structured in the following way.
\hyperref[sec:methodology]{Section~\ref{sec:methodology}} describes the research
methodology used for this research before then moving onto
\hyperref[sec:sec:evaluation]{Section~\ref{sec:evaluation}} which presents the
experimental results derived from a series of experiments conducted. This paper
then concludes with concluding remarks in
\hyperref[sec:conclusion]{Section~\ref{sec:conclusion}} where we discuss our
findings, detail the implications and propose potential future research avenues.

\begin{figure*}[t]
	\centering
	\includegraphics[width=\textwidth]{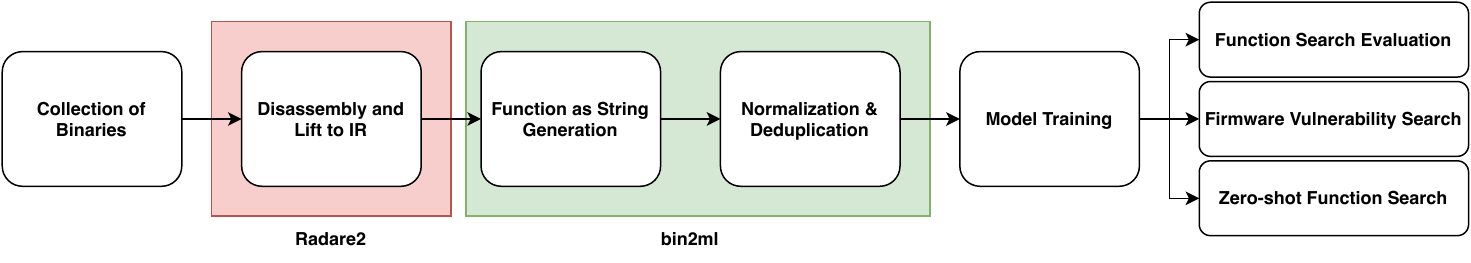}
	\caption{High Level Overview of Methodology}
\end{figure*}

\section{Methodology} 
\label{sec:methodology}

In this section we provide an overview of the methodology used to create our
experimental dataset and details related to how we train and evaluate our
proposed solution. We begin by describing our chosen IR before describing the
dataset used. We then move onto describe the process of going from raw binaries
to pre-processed, training-ready data. This section then continues to describe
the model design, training configuration and evaluation design before detailing
the metrics and baseline approaches used to compare against.

\subsection{Chosen Intermediate Representation}

The chosen IR is radare2's Evaluable Strings Intermediate Language (ESIL).
radare2 converts assembly language into a semantic equalivant, ESIL which
represents the architecture specific instructions using a combination of symbols
and numbers. \hyperref[fig:asm2esil]{Figure~\ref{fig:asm2esil}} provides several
examples of x86-64 assembly instructions and the corresponding ESIL
representations. The primary reason for choosing ESIL over other IR's such as
VEX, LLVM or PCode was compactness. Any given assembly instruction corresponds
to a single ESIL string. Whilst some assembly instructions create very large
ESIL string representations, through our experimentation the length of the ESIL
IR is typically shorter and more succinct as opposed to the alternatives.

\begin{figure}[htbp]
	\begin{center}
		\begin{BVerbatim}
	disasm: push rbp esil: rbp,8,rsp,0,=[8],8,rsp,-=
	
	disasm: call sym.imp.printf esil: 4176,rip,8,rsp,-=,rsp,=[8],rip,=
	
	disasm: mov dword [rbp - 8], 0 esil: 0,0x8,rbp,-,=[4]
		\end{BVerbatim}
		\caption{Example X86-64 and ESIL Representations \label{fig:asm2esil}}
		\end{center}
	\end{figure}

\subsection{Dataset}

In order to evaluate FASER and compare against comparative baselines, we use two
of the datasets detailed within~\cite{marcelli2022}. The first dataset is
\emph{Dataset-1} created by~\cite{marcelli2022}. \emph{Dataset-1} contains seven
popular open source projects: \texttt{ClamAv}, \texttt{Curl}, \texttt{nmap},
\texttt{Openssl}, \texttt{Unrar}, \texttt{z3} and \texttt{zlib}. Each of these
are compiled for ARM32, ARM64, MIPS32, MIPS64, x86 and x86-64 using four
different versions of Clang and GCC alongside 5 different optimization levels.
This results in each of the 7 projects having 24 unique compiler, architecture
and optimization combinations for each binary within the library.
Within~\cite{marcelli2022}, the authors formulate 6 tasks using this dataset
which increase in difficultly. For the purposes of this paper, we have chosen
the most difficult, denoted as \emph{XM}. The \emph{XM} task imposes no
constraints on which functions can be sampled from the corpus during test time
and includes all possible compiler, architecture and bitness combinations. This
task is representative of conducting binary function search against real
binaries.

The second dataset we use is the \emph{Dataset-Vulnerability} dataset also part
of~\cite{marcelli2022}. This dataset consists of two firmware images that
include several OpenSSL CVE's, specifically within the \emph{libcrypto} library
included as part of the firmware. The first firmware image is of a Netgear R7000
router which is ARM32 and a TP-Link Deco-M4 mesh router which is MIPS32. The
dataset also includes the same vulnerable library compiled for ARM32, MIPS32,
x86 and x86-64. The goal is to use the vulnerable functions from our compiled
libraries as a query function and then identify the corresponding vulnerable
function within the firmware image. In addition to the two tasks above, we
augment the second dataset with \emph{libcrypto} compiled for RISC-V 32-bit and
then re-run the firmware search. This task has been introduced to explore
whether an IR model is capable of transferring its learning to architectures it
has not seen before and can be considered a research first.

\subsection{Data Generation}

In order to generate the training data, \textit{bin2ml}~\cite{collyer2023bin2ml}
was used for both data extraction and pre-processing. \textit{bin2ml} uses
radare2 to disassemble the binaries and lift functions into ESIL IR.\@ Once this
lifting process has been complete, the data is then processed further to create
ESIL function as string representations by concatenating all ESIL instructions
for a given function into a single, long string. Any strings that were longer
than our model's input dimension were truncated. This could potentially cause a
loss of key information but is mitigated by the large input dimension chosen.
Each function string is then normalized using a series of heuristics before the
entire corpus is deduplicated. This process was repeated for all binaries within
our datasets. The specifics of the normalization and deduplication process are
presented below.

\subsection{Normalization}

Normalization is a fundamental step to ensure that the vocabulary size is
manageable, and all possible inputs can be encoded. In order to facilitate this,
a series of heuristics were applied to replace parts of the ESIL strings.  The
normalization approach draws on common approaches outlined within the literature
such as those used within SAFE \cite{massarelli2019safe}, jTrans
\cite{wang2022jtrans} and PalmTree \cite{li2021palmtree}. Firstly, any
hexadecimal value which starts with \texttt{0xfffff} or is one to three
characters long (such as \texttt{0x023} or \texttt{0x02}) are considered
immediate constants and replaced with the \emph{IMM} token. Secondly, any
hexadecimal value which starts with \texttt{0x} preceded by 4 or more
hexadecimal values is considered a memory address and replaced with \emph{MEM}.
Thirdly, due to the way radare2 represents function calls and data accesses,
these values are typically represented as integers within ESIL representations.
For this reason, if the opcode the ESIL representation was derived from is a
call opcode, the integer is replaced with the \emph{FUNC} token and otherwise
\emph{DATA}. And lastly, general purpose registers are replaced with tokens
based on their size, 32-bit registers are replaced with \emph{reg32} and 64-bit
registers are replaced with \emph{reg64}. As part of the experimentation, two
versions of FASER were trained, one without register normalization and one with.
This allows us to understand the impact register normalization would have on an
IR based model.

\subsection{Deduplication}

After normalization, deduplication takes place. Deduplication is critical
because even after changing factors such as optimization level and compiler, it
is still possible for binaries generated from the same source code to produce
identical functions. This is typically overlooked in existing literature and is
comparable to the approach presented in~\cite{marcelli2022}. For each of the
normalised ESIL strings, the ESIL string plus the function name are concatenated
into a single string before being hashed. These hashes are then compared with
each other to identify where there are duplicate functions. For any matches
found, only one was kept ensuring that the dataset used for training contains
only unique function strings. There is then a subsequent step taken which looks
through the entire dataset and eliminates functions which are only present once.
Essentially removing the functions whereby regardless of architecture or
optimization, after disassembly and lifting IR are invariant. Across our
dataset, the deduplication process eliminated on average 20-25\% of the
functions from a given library.

\begin{figure*}[htbp]
	\centering
	\begin{subfigure}[b]{0.49\textwidth}
		\centering
		\includegraphics[width=\textwidth]{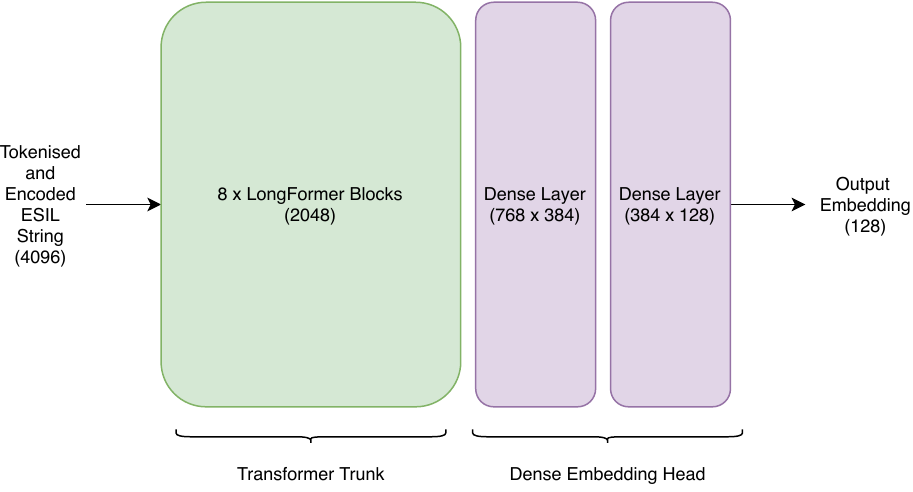}
		\caption{Overview of the Model Architecture Used}
		\label{fig:model_arch}
	\end{subfigure}
	\hfill
	\begin{subfigure}[b]{0.49\textwidth}
		\centering
		\includegraphics[width=\textwidth]{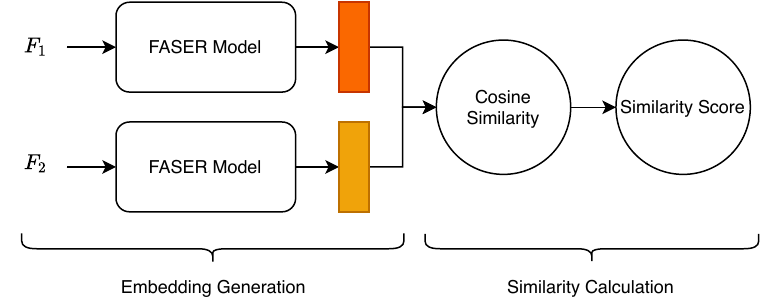}
		\caption{Siamese Training Formulation}
		\label{fig:model_training_setup}
	\end{subfigure}
\end{figure*}

\subsection{Model Design}

The chosen model used in FASER is a LongFormer. The LongFormer model was
proposed by~\cite{beltagy2020longformer} to tackle the quadratic computational
scaling of self-attention used in models such as BERT\@. The LongFormer instead
uses a combination of local, sliding window attention, with a global attention
mechanism. This formulation instead scales linearly as the input size increases,
providing a mechanism to train transformers with larger input sequences.
Furthermore, this combination of local and global attention is viewed favorably
for the binary function search task. An assembly instruction is not executed in
isolation but instead executed as part of a series of instructions. This local
attention window provides a means for a single instruction to include the
context of the instructions before and after it but in a manner which is
bounded. The global attention can then look at the function holistically whilst
being informed by the local contexts provided by the sliding window attention.

The model parameters used for FASER were an input dimension of 4096, followed by
8 LongFormer blocks with an intermediate dimension of 2048, followed by two
dense layers which map the 786 dimension transformer output to a 128 dimension
embedding. The local attention window is set to 512 tokens. We utilize the
implementation provided by \texttt{transformers} and all other parameters are
kept default. These can be view here~\cite{hugging2020longformer}

\subsection{Training Configuration}

Previous works such as Trex~\cite{pei2020trex}, jTrans~\cite{wang2022jtrans} and
PalmTree~\cite{li2021palmtree} conduct a pre-training step prior to then
fine-tuning for the function search objective. Whilst this makes sense if you
want to train a model which can be used for various different downstream tasks,
it's potentially suboptimal if the only downstream use case is going to be
function search. To this end, we forgo any pre-training steps and train FASER
directly for the function search objective using deep metric learning. We
construct a pair-based training methodology by using a Siamese formulation in
combination with Circle Loss \cite{sun2020circle}. Circle Loss was chosen due to
its ability to place emphasis on large deviations in between-class similarity in
a manner not possible with other losses such as triplet loss. Both Cosine
Embedding Loss and Triplet Loss were experimented with and resulted in unstable
training and in some cases, complete model collapse.

We also formulate a sampling strategy that ensures $m$ number of examples for a
given label (function) are present within a batch. We then apply the online
batch hard pair mining method \cite{hermans2017defense} to dynamically create
both positive and negative pairs for each example from batched inputs throughout
training. This works by embedding all examples within a batch before using the
associated labels for each example to create the hardest positive and negative
pairs for each example. What determines hardest is the output of a distance
function which in our case was Cosine Similarity. The strength of this approach
when compared to previous research which uses static pre-computed pairs is the
weaknesses of the model are consistently challenged. For example, if during
training the model quickly learns to search across ARM and MIPS but is
performing badly when searching across X86-64, this training formulation would
automatically begin to target this weakness by generating pairs including X86-64
examples and uses them to calculate the loss.

For training, we use the whole of Dataset-1 and sample 100K functions per epoch
for 18 epochs (Approximately 3 days of training). We set $m$ to 2 to ensure each
batch has 2 of each sampled function, batch size is set to 8 and, we use
gradient accumulation to artificially set the batch size to 512. The Adam
\cite{kingma2017adam} optimizer is used with a fixed learning rate of
\emph{0.0005}.

\subsection{Comparison Approaches}

For the first task, we draw upon the top performing approaches reported within
Marcelli (2022)\cite{marcelli2022} which are the Graph Matching Networks (GMN)
and Graph Neural Network (GNN) approach from Li et al (2019)\cite{li2019graph}
and Gemini \cite{xu2017neural}. All of these are Graph Neural Network (GNN)
approaches which take advantage of the structural aspects of functions,
typically through using the control flow graph (CFG) with node level feature
vectors as an input. Approaches using natural language processing and
transformer model architectures such as PalmTree \cite{li2021palmtree} and
jTrans \cite{wang2022jtrans} would have been ideal candidates but are
mono-architecture therefore were deemed unsuitable for comparison. 

For the vulnerability search task, we use the same three approaches outlined
above but also compare against Trex \cite{pei2020trex}. Trex provides an
interesting comparison because it too uses a transformer architecture but has
one significant difference. The model is pre-trained on what the authors
describe as micro-traces. These micro-traces are generated in a dynamic manner
using an emulator. Once trained, the model is capable of being used with solely
static data and forgoes the emulation aspect. The emulator used to generate the
micro-traces provided by the authors does not support the full breadth of
architectures and bitness of Dataset-1 therefore would be an unfair comparison
and therefore, is not used in the first task.

\subsection{Evaluation Configuration}

For task 1, we again use Dataset-1 and implement a sampling approach which
dynamically creates \emph{search pools} which for a given function, contain 1
positive example and 100 negative examples. This formulation is the same
as~\cite{marcelli2022}. We also adopt the same methodology
as~\cite{marcelli2022} for the Vulnerability search task whereby we have a query
function of a given architecture and search across all possible functions in the
firmware's \textit{libcryto} library. This means that the search pool size for
task 2 is 10 times bigger at approximately 1000 functions.

\subsection{Metrics}

We re-use the metrics used in previous studies \textbf{Recall@1} and
\textbf{MRR@10} for task 1 in order to present a reliable comparison. For the
vulnerability search task, we report the rank at which the vulnerable function
was present at after the search was conducted similar to other studies,
alongside this, we also calculate the mean and median ranks across all
architecture searches. This is primarily to aid result analysis.

\section{Evaluation}
\label{sec:evaluation}

Our evaluation aims to answer the following research questions:
\begin{enumerate}
	\item \textbf{RQ1} - How does FASER perform when compared against other
	baseline approaches for the binary function search task?
	\item \textbf{RQ2} - How effective is FASER at searching real firmware
	images for known vulnerabilities?
	\item \textbf{RQ3} - Does using intermediate representations as the input
	data enable the model to zero shot architectures not previously seen as part
	of the training data?
\end{enumerate}

\subsection{RQ1 - Binary Function Similarity Search}

The results of the experimentation to gather data for \emph{RQ1} can be seen in
\hyperref[table:rq1-results]{Table 1}. Both of the FASER models trained
outperform all the baseline approaches across both of the chosen metrics.
Looking first at Recall@1, the model without register normalized training data
(denoted as FASER NRM) performs significantly better than the register
normalized model, achieving a Recall@1 increase of 13\% when compared against
the best performing baseline, GMN\@. FASER RN performs comparable to the GMN
model without needing direct comparison between all possible function
combinations within a given search pool. 

Moving onto the MRR@10 results, FASER NRM again performance significantly better
than all baselines with a 7.5\% increase. FASER RM is again comparable to the
GMN approach with an identical MRR@10.

\begin{table}[htbp]
\small\selectfont
\begin{tabular}{llrr} \toprule &                      &
                             					\multicolumn{2}{c}{\textbf{XM}}
                             					\\ \midrule
                             					\textbf{Method}
                             					& \textbf{Description} &
                             					\textbf{R@1}  & \textbf{MRR@10}
                             					\\\midrule FASER NRM & ESIL
                             					Function String & \textbf{0.51}
                             					&  \textbf{0.57} \\
FASER RN               					        & ESIL Function String & 0.46 &
0.53 \\
GMN \cite{li2019graph} 						  	& CFG + BoW opc 200    & 0.45 &
0.53            \\
GNN \cite{li2019graph}                          & CFG + BoW opc 200    & 0.44 &
0.52            \\ 
GNN (s2v) \cite{xu2017neural}                   & CFG + BoW opc 200    & 0.26 &
0.36            \\ \bottomrule
\end{tabular}
\centering
\caption{RQ1 - Binary Function Similarity Search. NRM = No Register
normalization and RN = Register normalization\label{table:rq1-results}}
\end{table}

\textbf{RQ1 Summary:} The results present above show that our propose approach,
FASER, performs as well as if not better than the best baseline approach with
FASER NRM performing better across all metrics.

\subsection{RQ2 - Binary Function Vulnerability Search}

The results of the experimentation to gather data for \emph{RQ2} can be seen in
\hyperref[table:rq2-results]{Table 2}. The results show the ranks of search
results when searching the Netgear R7000 router which is ARM32. 

\begin{table*}[!htbp]
	\centering
	\small{
	\begin{tabular}{@{}lrrrrrr@{}}
	\toprule
	\textbf{}                              
	& \multicolumn{6}{c}{\textbf{NETGEAR R700}} \\ \midrule & \textbf{ARM} &
	\textbf{MIPS} & \textbf{X86} & \textbf{X86-64} & \textbf{Mean} &
	\textbf{Median} \\
	&&&&& \textbf{Rank} & \textbf{Rank} \\\midrule \textbf{GNN} & 4:1:1:44     &
	97:5:1:138 & 3:31:1:18    & 9:6:1:40        & 25 & 5.5 \\
	\textbf{GNN (s2v)}                     & 2:1:1:6      & 35:5:1:7 & 8:1:5:36
	& 9:1:14:8        & 8.125              & 5.5 \\
	\textbf{Trex}                          & 32:4:1:2     & 24:16:1:1 & 41:4:1:3
	& 10:3:1:2        & 9.125              & 3 \\
	\textbf{GMN}                           & 1:1:1:1      & 1:1:1:7 & 1:1:1:2 &
	1:1:30:7        & 3.625              & 1 \\ \midrule \textbf{FASER NRM} &
	1:5:1:1      & 9:122:1:3     & 21:21:3:50   & 7:122:1:2       & 23.125 & 4
	\\
	\textbf{FASER RN}  			   		   & 1:6:1:1      & 2:4:1:4 & 2:1:1:3 &
	1:2:1:1         & 2                  & 1 \\ \bottomrule

	\end{tabular}%
	}
	
	\caption{RQ2 - Vulnerability Search Result
	Rankings\label{table:rq2-results}}
\end{table*}

In addition to the three baseline approaches used in the previous task, our
proposed approach was compared against Trex, a comparable transformer based
approach which has a more complicated and computationally expensive training
process. The results presented in \hyperref[table:rq2-results]{Table 2} show
that our proposed approached performs well across all the architectures.
Interestingly, whilst the register normalized model performed the strongest in
the Binary Function Similarity Search task, in this task the register normalized
model performs significantly better. This is shown by both mean and median rank
descriptive statistics being lower. The best performing FASER model is highly
comparable to the GMN method but again without the aforementioned limitations.
Comparing specifically to Trex, the register normalized model performs
consistently better across all the architectures. This suggests that our
training methodology of training for the function similarity directly and
forgoing the elaborate pre-training steps usually adopted and the use of IR's as
our data input has merit.

\textbf{RQ2 Summary:} The results present above show that the FASER RN performs
well when searching real firmware images for known vulnerabilities.

\subsection{RQ3 - Zero Shot Architecture Binary Function Search}

\hyperref[table:rq3-results]%
{Table~\ref{table:rq3-results}} shows the results from the experimentation
undertaken to answer \emph{RQ3}. The question posed here is can the FASER
models, because we are using an IR as the input data, perform zero-shot
vulnerability search for a new architecture by transferring prior learnt
knowledge. Fundamentally, the answer to this is no. The vulnerability search
performance for a new instruction set architecture (in this case RISC-V) is
significantly worse. This is clearly demonstrated by the mean and median rank
descriptive statistics.

Nevertheless, these results do demonstrate something interesting. Across both
FASER models, the performance is significantly better when searching MIPS
functions using a RISC-V query as opposed to searching ARM functions using a
RISC-V query. This suggests that the semantic representation created when MIPS
and RISC-V instructions are lifted to ESIL may be more similar than ARM and
RISC-V. Given that recent research has suggested that ARM and X86/X86-64
instructions are closer in statistical similarity than when compared to MIPS
\cite{kim2022tiknib}, these results may suggest that introducing RISC-V binaries
into a training dataset may level out any data imbalances.

\begin{table*}[h]
	\label{table:rq3-results}
	\centering
%	\resizebox{\textwidth}{!}%
	{\small%
	\begin{tabular}{llrrlrr}
	\toprule
	& \textbf{ARM32}  & \textbf{Mean} & \textbf{Median} & \textbf{MIPS32}   &
\textbf{Mean} & \textbf{Median} \\
&   & \textbf{Rank} & \textbf{Rank} & & \textbf{Rank} & \textbf{Rank} \\
   \midrule FASER NRM & 48;546;964;14   & 393                & 297 &
   48;30;22;546;170; & 154                & 101.5 \\
	 &    & 393                & 297 & 251;14;155 & 154                & 101.5
	\\
	FASER RM  & 673;292;1004;15 & 496                & 482.5 & 673;33;4;292;76;
	& 172                & 106  \\ 
	&& 496                & 482.5 & 136;15;147  & 172                & 106
	\\\bottomrule             
	\end{tabular}%
	} \caption{RQ3 - Zero Shot Architecture Search\label{table:rq3-results}}
	\end{table*}

\textbf{RQ3 Summary:} The results present above show that the use of an IR input
representation does not provide a means of conducting zero-shot search across
unseen architectures.

\section{Discussion and Conclusion}
\label{sec:results}

\label{sec:conclusion}

The results presented above demonstrate that the combination of ESIL IR and the
Longformer transformer architecture perform well compared to the baseline
approaches with minimal requirement for either manual feature engineering or
dynamic analysis. The FASER RN model performs particularly well at the
vulnerability search task across all architectures tested and performs
comparably to GMN, without requiring direct comparison between all possible
combinations within a given search pool. Whilst demonstrating the effectiveness
of IRs and longer context transformers, these results also add weight to our
argument that the pre-training step seen within previous work may be unwarranted
for binary function search and training for the binary function similarity
objective directly may be more optimal.

The results presented to answer RQ3 also suggest something interesting. While
the search rank results were significantly worse in terms of mean and median
rank, suggesting that our proposed approach is unable to reliably transfer to
unseen architectures, there is a large difference between the reported ranks for
RISC-V $\rightarrow$ ARM when compared to RISC-V $\rightarrow$ MIPS ranks. Prior
research \cite{kim2022tiknib} observed similar phenomena whereby x86
$\rightarrow$ ARM functions were statistically more similar than x86
$\rightarrow$ MIPS functions. This suggests that RISC-V functions may be more
similar to MIPS functions in terms of semantics when represented in ESIL than
ARM\@. Given that most datasets used in prior research only include X86, ARM and
MIPS, this similarity could potentially be leveraged and experimented with
further. An example of this experimentation could be to see explore whether
including RISC-V functions within a cross-architecture binary function search
dataset balances out the x86 $\rightarrow$ ARM similarities with providing
function examples that are similar but different to the MIPS architecture. 

Turning now to implications of this research. Firstly, this work demonstrates
that IRs derived from binaries can be used to train models for binary function
search and perform well. Secondly, the use of longer input sequences also works
well. The performance results, especially related to the use of the LongFormer
architecture suggest that changing the type of transformer architecture used and
increasing the input dimension for approaches such as jTrans
\cite{wang2022jtrans}, Trex \cite{pei2020trex} or PalmTree \cite{li2021palmtree}
may increase their overall performance. The results also demonstrate that  if
the only downstream target task is binary function search it may be worth
amending the standard training methodology which involves a pre-training step
and instead, train for the objective directly.

In future work, there are several avenues that could be explored. There are a
number of different IRs that could be incorporated into similar approaches such
as VEX \cite{shoshitaishvili2015firmalice}, LLVM \cite{LLVM:CGO04} and PCode
\cite{nsa2019ghidra}. There is also an emerging sub-field of binary function
search focused on adding heuristic pre- and post-filtering steps to increase
performance by reducing the number of functions searched such as those described
in Asteria-Pro \cite{yang2023asteria} and BinUSE \cite{wang2022enhancing}. And
finally, this approach could be enhanced through the integration of supporting
models such as those that use decompiled source code, recovered type information
or structural aspects at a control flow graph or call graph level. 

%\printbibliography
\bibliography{lit}
\bibliographystyle{plain}
\end{document}